\documentstyle[epsfig,12pt]{article}
\def\epem{$e^+ e^-$}

\begin{document}

\title{ A nonextensive thermodynamical equilibrium \\ 
approach in $e^+ e^- \to hadrons$}

\author{I. Bediaga, E. M. F. Curado and J.M. de Miranda \\ 
Centro Brasileiro de Pesquisas F\'\i sicas, \\ 
Rua Xavier Sigaud 150, 22290-180 -- Rio de Janeiro, RJ, Brazil }

\maketitle

\begin{abstract}

	We show that the inclusion of long distance effect, expected in 
strong interactions, through a nonextensive thermodynamical approach 
is able to explain the experimental distribution of the transverse 
momentum of the hadrons with respect to the jet axis ($p_{t}$) 
$e^+ e^- \to hadron$ reaction. 
The observed deviation from the 
exponential behavior, predicted 
by the Boltzmann-Gibbs thermodynamical treatment, is automatically recovered 
by the nonextensive Tsallis statistics used here. We fitted the 
observed $p_{t}$ spectrum in the range of 14 GeV to 161 GeV 
and obtained, 
besides a good fit, the theoretical important fact that the
temperature  becomes independent of the 
primary energy.

\end{abstract}

\vspace{3cm}

\begin{tabbing}

\=xxxxxxxxxxxxxxxxxx\= \kill

\>{\bf Keywords:} \> High Energy; Hadroproduction; Fireball; Statistical Models. \\

\>{\bf PACS Numbers:} \> 24.10.Pa, 13.65.+i, 13.60.Hb,  

\end{tabbing}

\newpage

\section{Introduction}\label{sec.intro}

\indent

The global structure of multiple hadroproduction in \epem annihilation, 
has been well understood
with remarkable results concerning the total cross section and angular 
distribution calculations. Perturbative Quantum Electrodynamics and 
perturbative Quantum Chromodynamics provide a good description of the 
initial process involving short distances $e^+e^-\to$ quark-antiquark 
($q\bar q$) and $e^+e^-\to$ quark-antiquark-gluon ($q\bar qg$) 
interactions, main responsible for the global characteristics of this kind 
of process. The quarks produced off-shell initialize a cascade of several 
quarks and gluons with complex interactions themselves. This stage is in 
general described in probabilistic terms, based on leading-log Quantum 
Chromodynamics approximation. The outgoing colored partons are transformed 
into color singlet hadrons through the soft hadronization process, forming 
a jet of particles traveling approximately in the initial parton 
direction. However, hadronization require the Quantum Chromodynamics in 
the soft regime, where the coupling constant become large and 
there is no easy way to understand it from first principles.

	For more than a decade several different models making use of Monte Carlo 
techniques were developed to describe such processes 
\cite{herwig,jetset,ucla}. These methods avoid the microscopic complexity
by being able to define probability distributions for each stage of the jet
evolution. In these approaches, all known aspects involving Quantum 
Chromodynamics were included, the unknown aspects like hadronization being 
modeled phenomenologically. Each model has a set of free parameters which 
enables it to reproduce many distributions over a
wide range of energy. In particular, the transverse momenta of the 
hadrons with respect to the jet direction, main subject of this paper, 
has been accurately parameterized. Nevertheless, there is no fundamental 
understanding of the matter \cite{ucla}.

	One alternative approach to understand the transverse
 momentum distribution of the charged hadrons produced in 
 \epem  annihilation 
 is to look at the whole process in a macroscopic way, trying to extract 
 information through a thermodynamical treatment, not taking into  
 account the microscopy interactions between partons, governed  by the 
 Quantum Chromodynamics \footnote{We used the total transverse momentum 
 instead of $p_{t}^{in}$ and $p_{t}^{out}$  
because we are looking at the global characteristics of the events }.  
Fermi \cite{fermi} was the pioneer in this 
 kind of approach, followed by Hagedorn \cite{hagedorn} who provided 
 the most consistent and well succeeded model. Nowadays, the thermodynamical 
 formalism is widely used in heavy ions collisions \cite{heinz}
 and has also been applied in $e^+ e^-$,
proton-proton ($pp$) and proton-antiproton ($p\bar p$) interactions to 
determine the multiplicity of hadrons produced, with noteworthy results 
\cite{becattini1,becattini2,becattini3}.

\section{Statistical approach}\label{sec.statShannon}

\indent 

According to Hagedorn, a high energy collision produces an 
\emph{unlimited} and \emph{undetermined} number of different excited 
hadron \emph{fireballs} that reach a thermodynamical equilibrium. In this 
relativistic statistical \emph{Ansatz}, not only the number of particles 
but also the number of available types of
particles itself grows with energy. The immediate and important 
consequence of this production mechanism is that \emph{the temperature is
independent of the primary energy} \cite{hagedorn}. A similar statement was
also made, a few years later, by Field and Feynman \cite{feynman}. One 
should expect that this temperature governs the transverse momentum 
($p_{t}$) distribution of the outgoing particles with respect to the  
jet axis. 

In a modern Ansatz \cite{becattini1}, one can assume that the 
thermodynamical equilibrium is reached in each jet produced in 
$e^+ e^- \to hadrons$ interaction.  We consider in this paper 
that the  thermalization  
begins just after the hadronization process takes place and finishes when 
the hadrons are no longer interacting. Since the range of this 
interaction is of order of a few Fermi, the resonances produced, 
with a typical lifetime of order of $10^{-23}$ s, should 
decay while the particles are still interacting.   
Thus, the particles  
produced from these decays should also participate in the thermalization
process.  As most charged particles observed in the detector  
are produced prompt or through the decay of hadronic resonances, we could  
expect that the charged particles observed in the detectors have their 
distributions dictated by the previous `thermodynamical' equilibrium.

It is then important to determine the thermodynamical properties of the event, 
without any kinematical effect caused by the fast relative motion between 
the hadrons. Using the Boltzmann-Gibbs statistics in this large grand-canonical 
ensemble for a relativistic gas, the transverse momentum distribution 
with temperature $T_{0}$, can be written as (see Hagedorn \cite{hagedorn}): 

\begin{equation}
\frac{1}{\sigma} \frac{d\sigma}{dp_{t}} = cp_{t} \int_{0}^{\infty} dp_{\it 
l} \exp{(-
\frac{1}{T_{0}}\sqrt{p_{\it l}^{2}+\mu^{2}})} \hspace{0.1cm},
\label{eq:pl}
\end{equation}
where $p_{\it l}$ means the longitudinal momentum, $\mu^{2} \equiv 
p_{t}^{2} + m^{2}$
and $m$ is the mass ($m\ll p_{t}$) \cite{hagedorn}. After the integration 
indicated in eq. (\ref{eq:pl}) and for $p_{t}/T_{0} \gg 1 $ we obtain an 
exponential asymptotic distribution for the transverse momentum 
\cite{hagedorn,heinz}:

\begin{equation}
\frac{1}{\sigma} \frac{d\sigma}{dp_{t}} \approx c p_t^{\frac{3}{2}} 
\exp{(-\frac{p_{t}}{T_0})}
\hskip 0.1 cm .
\label{eq:ptb}
\end{equation}

\noindent
As the temperature should not depend on the center-of-mass energy 
($E_{CM}$),
we can expect that eq. (\ref{eq:ptb}) presents the same shape for all $e^+ 
e^-$ energies spectrum varying
only the parameter $c$, which depends on the average multiplicity of the 
events and consequently on the $E_{CM}$. 

The measured transverse momentum distributions of charged hadrons 
with respect to the jet axis, defined as the sphericity 
axis  \cite{tasso,delphi}, for 
center-of-mass energies in the range of 14 GeV to 161 GeV are displayed in 
figure 1. One notices the clear experimental deviation, for high values of 
$p_{t}$, from the exponential behavior predicted by eq. (\ref{eq:ptb}) 
and indicated by a dotted line. Notice that we could also get decent fits 
for low values of data point of $p_{t}$ ($p_{t}<1.5$ GeV) 
using the eq. (\ref{eq:ptb}), but in this case the best values for the 
parameter $T_0$ would clearly increase with energy, a feature which 
disagrees with Hagedorn's physically reasonable assumption.

\begin{figure}[hbt]
\centerline{\epsfig{file = 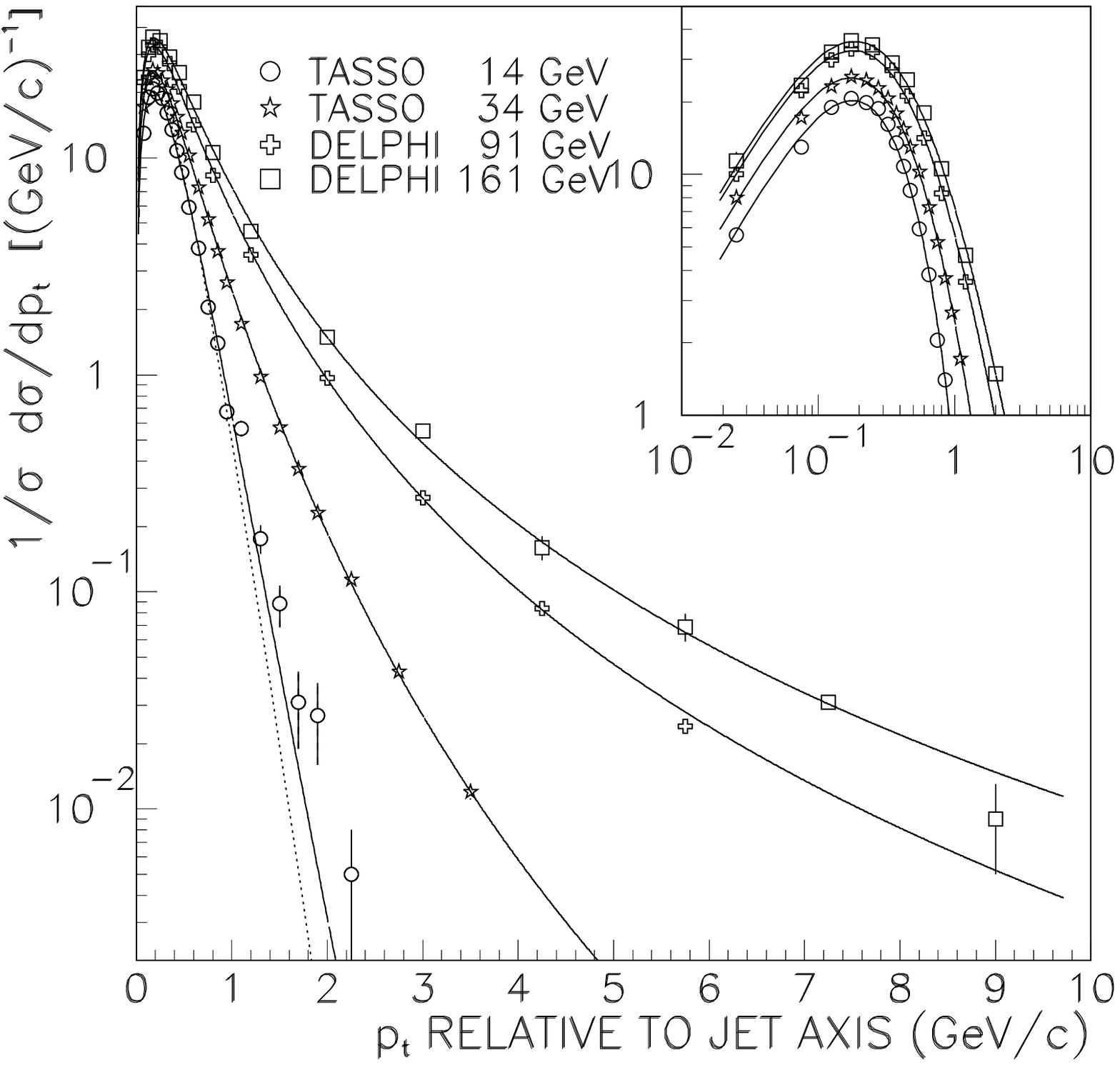,width=8cm}} \caption{Transverse 
momentum distribution . The distribution $\frac{1}{\sigma} 
\frac{d\sigma}{dp_{t}}$ of the transverse momentum $p_{t}$ of charged
hadrons with respect to jet axis (defined in these experimental 
results as the sphericity axis) is sketched for 
four different experiments, whose center of mass energies vary from 14 and 
34 Gev (TASSO) up to 91 and 161 Gev (DELPHI). The Hagedorn predicted 
exponential behaviour is shown by the dotted line. We
can see that the deviation of the exponential behaviour increases when the
energy increases.
The continuous lines are obtained from our eq. (\ref{eq:pt}) and agree 
very-well with the experimental data. The inset shows the transverse 
momentum
distribution for small values of $p_{t}$.} \label{fig1} \end{figure}

\section{Alternative statistical approach}\label{sec.statT}

\indent 

The Boltzmann-Gibbs statistics, used to derive eq. (\ref{eq:ptb}), 
is suitable for an enormous number of systems, but it is known that 
it could lead to unphysical results for some systems that 
present long-range interactions (LR) 
\footnote{Note that we are using the expression - long range interactions - 
in the statistical mechanical sense, meaning interaction among particles 
that are far away one of the other. This situation include both cases, 
soft and hard interactions in QCD.}.  
In fact, LR interaction is the usual interpretation for the 
intermittency phenomena \cite{kittel}, large particle density  
fluctuations \cite{NA22,JACEE}, present in hadroproduction. 
Long-range (LR) interactions can occur in many situations, 
particularly in  
processes where the interacting volume is of order of the  
dimension of the hadrons themselves, resulting in a high hadronic 
density.
In this way the gradual deviation of the 
observed distribution from the exponential function, shown in figure 1, 
could be interpreted as a gradual increase of the 
multiplicity and the 
hadronic density in the  thermalization mechanism. 

Therefore it is useful, to look for another 
statistics that, unlike the Boltzmann-Gibbs, could treat these 
systems. Recently
Tsallis \cite{tsallis} proposed a generalized statistics whose formalism 
\cite{tsallis,termo} has been able to describe many systems
having as common features the presence of LR interactions and/or 
non-Markovian processes \cite{polytropic}. This formalism is based on 
the entropic form $S_{q} = (1 - Tr \rho^q)/(q-1)$ 
where $\rho $ is the density matrix (state) of the system and $q$ is a 
real positive parameter.  When $q \rightarrow 1$ the above entropy tends 
to the well-known Boltzmann-Gibbs-Shannon (BGS) entropy 
$S =  - Tr \rho \log \rho$. 
In this sense the $S_{q}$ entropy extends the BGS entropy to the nonextensive 
cases corresponding to $q \neq 1$. In fact, in this case the additivity 
property of the entropy fails because it appears a new term proportional 
to $q-1$. For example, if we look the way the composed entropy of 
two sub-systems $A$ and $B$,  
$S_{A + B}$, is related to the entropy of each of the two sub-systems, 
$S_{A}$ and $S_{B}$, when the sub-systems $A$ and $B$ are statistically 
independents, 
one gets the expression  
$S_{A+B} = S_{A} + S_{B} + (1-q)S_{A} S_{B}$. It is easy here to see 
that the case $q=1$ reobtains the standard result (additivity) 
from Boltzmann-Gibbs entropy \cite{termo,tsallisrev}.    
The extremization of the $S_{q}$ entropy with  
adequate constraints lead us to a probability distribution that exhibits a 
power-law decay (instead of the exponential decay presented by the BGS 
entropy).  Examples of application of this statistics are available in 
several papers \cite{polytropic}, whose studied systems are, in general, 
nonextensive. 
Thus, assuming a thermal equilibrium of a 
relativistic gas (the \emph{fireball}) that obeys the Tsallis statistics  
we obtain, observing the derivation 
made by \cite{turcos}, the following equation for the transverse 
momentum distribution (which replaces  eq. (\ref{eq:pl})):

\begin{equation}
\frac{1}{\sigma} \frac{d\sigma}{dp_{t}} = c p_{t}\int_{0}^{\infty} dp_{\it 
l}
[1 - \frac{1-q}{T_{0}}\sqrt{p_{\it l}^{2}+\mu^{2}}]^{\frac{q}{1-q}} 
\hspace{0.1cm},
\label{eq:pt}
\end{equation}

\noindent
where $q$ is related
to the degree of non-extensivity of the system 
\cite{tsallis,termo,tsallisrev}.
$T_{0}$, $p_{\it l}$, $c$ and $\mu$ have the same meaning as in the 
previous equations.
The Boltzmann-Gibbs limit - eq. (\ref{eq:pl}) - is achieved as $q \to1$. 
The exact expression of the transverse momentum distribution, eq. 
(\ref{eq:pt}), is written in the appendix.

\noindent
With this expression we fit the experimental data leaving $q$, $T_0$ 
and $c$ as free parameters and the best fits are presented in the figure 1. 
We can see that the asymptotic behaviour 
of the new distribution, for larger values of $p_{t}$,
represents very-well the deviation
from the exponential for \emph{all} center-of-mass energies. 

\begin{table} \centering
\begin{tabular}{|c|c|c|c|}	\hline
Experiment & Energy (GeV) & $q$ & $T_0$(GeV) \\ \hline \hline

TASSO & 14 &1.020 $\pm 0.005 $ & 0.130$ \pm 0.002 $ 
\\ \hline
TASSO & 34 &1.1225 $\pm 0.0006 $ &0.1152 $ \pm 0.0004 $ 
\\ \hline
DELPHI & 91 &1.1938 $\pm 0.0005 $ & 0.1094$ \pm 0.0004$ 
\\ \hline
DELPHI & 161&1.215 $\pm 0.002 $ & 0.110 $ \pm 0.002 $ 
\\ \hline
\end{tabular}
\protect\caption{}
\label{table1}
\end{table}

	 In figure 2 we present $q$ and $T_{0}$ values as functions of the \epem 
center-of-mass energy. Their numerical values are shown in table 1. We can 
see that the $q$ parameter increases smoothly with the energy, saturating 
around $1.2$. On the other hand, the temperature has a pretty constant 
behaviour, $T_{0} \approx 0.11$ GeV, for the higher energy range. We note 
that this constant value for the temperature is not an ad hoc hypothesis; 
rather, it was obtained by the fit, 
indicating that all the energy given to the \epem system 
is being used to create new 
\emph{fireballs}, as predicted by Hagedorn \cite{hagedorn}. 
We note that our $T_{0}$ value is lower than the value obtained 
in reference \cite{becattini1}, and this difference is probably due 
to the fact that, unlike us, the author considered the thermal equilibrium 
just after the hadronization, i.e., before the 
decay of the hadronic resonances. 
In our process, the interaction volume continue to expand, lowering the 
temperature, until the particles no longer interact   
among themselves.
Also, the statistics used in both papers are different. 
The smooth 
increase of $q$ with the center of mass energy can be understood as the
expected increase of the influence of the  
LR interactions among the hadronic particles produced in the event.

\begin{figure}[hbt]
\centerline{\epsfig{file = 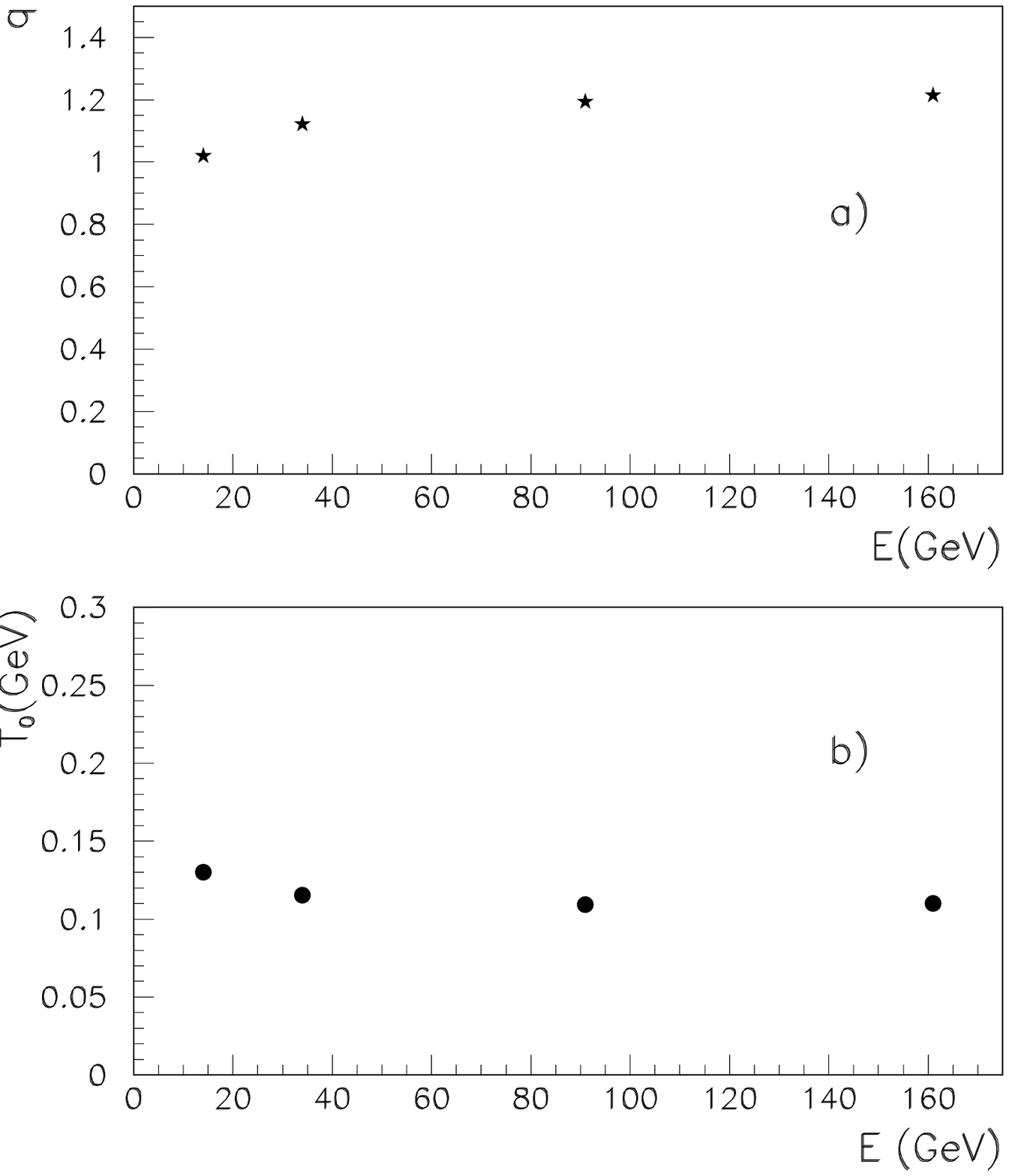,width=8cm}} \caption{Variation of 
(a) the entropic index ($q$) and (b) the temperature ($T_{0}$) with 
the center of mass energy. The entropic index $q$ (stars) increases 
slightly and the temperature (squares) remains almost 
constant, as predicted by Hagedorn, . The center-of-mass energies 
vary from 14 to 161 Gev. The error bars are smaller than 
the size of the symbols.}
\label{fig2}
\end{figure}

	The deviation from an exponential behaviour observed in the transverse 
momentum distribution in the \epem interaction, is also observed for 
hadrons produced in other interactions like $p \bar p$ or heavy ions. It 
indicates a universal characteristic of the dynamics of hadron production, 
probably due to the relevance of the LR regime,  
related to the high hadronic density. 

\section{Conclusion}\label{sec.conclusion}

\indent 

The thermodynamical treatment using the Tsallis statistics 
discussed here presented a remarkable agreement with the experimental data 
in more than 4 orders of magnitude in the differential cross section - 
emphasized with the predicted constant behavior of the temperature $T_{0}$.
Therefore, the results of this approach suggest an alternative statistical 
treatment for the transverse momentum distribution within the 
accepted theory of strong interaction. 

\vspace{1cm}

Acknowledgments. We thank Dr. Oliver Passon for detailed DELPHI 
experimental data and Carlos Aguiar for useful discussion. One of us, 
E.M.F.C., thanks Pronex/MCT for the  partial financial support. 

\hspace{0.5cm}

E-mail: bediaga@cbpf.br, eme@cbpf.br, jussara@lafex.cbpf.br

\newpage

\section*{Appendix }

After some manipulation, the solution of the integral in 
eq. (\ref{eq:pt}) was obtained by the software
Mathematica 3.0 and the transverse momentum distribution can 
be written as:

\begin{eqnarray*}
\begin{array}{l}
\frac{1}{\sigma}\frac{\partial\sigma}{\partial p_{t}} = (T^{2}\sqrt{\frac
{T}{T+(q-1)p_{t}}}(-(2^{3-\frac{q}{q-1}}\left( 
1+\frac{T}{(q-1)p_{t}}\right) ^{\frac{1}{2}+\frac{q}{q-1}}\left( 
\frac{q-1}{T}p_{t}\right) ^{3}\Gamma\left( \frac{q}{q-1}\right)* \\
\hspace{1.37cm} (4\Gamma\left( \frac{5}{2}-\frac{q}{q-1}\right) 
\Gamma\left( \frac {q}{q-1}-2\right)* \\
\hspace{1.37cm} _{2}F_{1}\left( 
\frac{q}{q-1}-2,\frac{q}{q-1},\frac{q}{q-1}% 
-\frac{3}{2},\frac{(q-1)p_{t}+T}{2(q-1)p_{t}}\right) + \\ 
\hspace{1.37cm} 2\Gamma\left( \frac{3}{2}-\frac{q}{q-1}\right) 
(\Gamma\left( \frac {q}{q-1}-2\right)* \\
\hspace{1.37cm} _{2}F_{1}\left( 
\frac{q}{q-1}-2,\frac{q}{q-1},\frac{q}{q-1}% 
-\frac{1}{2},\frac{(q-1)p_{t}+T}{2(q-1)p_{t}}\right) +\\ 
\hspace{1.37cm} \Gamma\left( \frac{q}{q-1}-1\right) \hspace{0.1cm} 
_{2}F_{1}\left( \frac{q}% 
{q-1}-1,\frac{q}{q-1},\frac{q}{q-1}-\frac{1}{2},\frac{(q-1)p_{t}+T}{2(q-1)p_{t}}\right) 
) + \\ 
\hspace{1.37cm} \Gamma\left( \frac{1}{2}-\frac{q}{q-1}\right) \Gamma\left( 
\frac{q}{q-1}-1\right)* \\
\hspace{1.37cm} _{2}F_{1}\left( 
\frac{q}{q-1}-1,\frac{q}{q-1},\frac{q}{q-1}% 
+\frac{1}{2},\frac{(q-1)p_{t}+T}{2(q-1)p_{t}}\right) )) - \\ 
\hspace{1.37cm} \sqrt{2}\left( 1+\frac{q-1}{T}p_{t}\right) \pi^{2} 
(8\left( \left( \frac{q-1}{T}\right) p_{t}\right)^{2}* \\ \hspace{1.37cm} 
_{2}F_{1}Regularized\left( -\frac{1}{2},\frac{1}% 
{2},\frac{3}{2}-\frac{q}{q-1},\frac{(q-1)p_{t}+T}{2(q-1)p_{t}}\right) + \\ 
\hspace{1.37cm} \left( 1-\left( \frac{q-1}{T}\right) p_{t}\right) 
(-4\left( \frac {q-1}{T}\right) p_{t}* \\
\hspace{1.37cm} _{2}F_{1}Regularized\left( -\frac{1}{2},\frac{3}% 
{2},\frac{5}{2}-\frac{q}{q-1},\frac{(q-1)p_{t}+T}{2(q-1)p_{t}}\right) + \\ 
\hspace{1.37cm} 2\left( \frac{q-1}{T}\right) p_{t} \hspace{0.04cm} 
_{2}F_{1}Regularized\left( 
\frac{1}{2},\frac{3}{2},\frac{5}{2}-\frac{q}{q-1}% 
,\frac{(q-1)p_{t}+T}{2(q-1)p_{t}}\right) - \\ 
\hspace{1.37cm} 3\left( 1-\left( \frac{q-1}{T}\right) p_{t}\right)* \\ 
\hspace{1.37cm} _{2}F_{1}Regularized\left( \frac{1}{2},\frac{5}% 
{2},\frac{7}{2}-\frac{q}{q-1},\frac{(q-1)p_{t}+T}{2(q-1)p_{t}}\right) 
))Sec\left( \frac{\pi q}{q-1}\right) )) / \\ \hspace{1.37cm} \left( 
16\sqrt{\left( \frac{q-1}{T}p_{t}\right) }\left( 1+\frac{q-1}% 
{T}p_{t}\right) ^{\frac{q}{q-1}}\sqrt{\pi}\Gamma\left( 
\frac{q}{q-1}\right) \left( q-1\right) ^{2}\right)
\end{array}
\end{eqnarray*}

\noindent
where $_{2}F_{1}(\alpha,\beta;\gamma;z)$ means the generalized 
hypergeometric series, \\ 
$_{2}F_{1}Regularized(\alpha,\beta;\gamma;z) \hspace{0.1cm} \equiv 
\hspace{0.1cm} _{2}F_{1}(\alpha,\beta;\gamma;z)/\Gamma(\gamma)$  and \\ 
$\Gamma(x)$ is the Gamma function.

\end{document}